# Does ESG and Digital Transformation affects Corporate Sustainability? The Moderating role of Green Innovation

**Chenglin Qing[1], Shanyue Jin[2]***

[1]Department of Business Administration, Honam University, South Korea

[2]College of Business, Gachon University, South Korea

**\* Correspondence:**
Shanyue Jin
jsyrena0923@gachon.ac.kr

## ABSTRACT

Recently, environmental, social, and governance (ESG) has become an important factor in companies' sustainable development. Artificial intelligence (AI) is also a core digital technology that can create innovative, sustainable, comprehensive, and resilient environments. ESG- and AI-based digital transformation is a relevant strategy for managing business value and sustainability in corporate green management operations. Therefore, this study examines how corporate sustainability relates to ESG- and AI-based digital transformation. Furthermore, it confirms the moderating effect of green innovation on the process of increasing sustainability. To achieve the purpose of this study, 359 data points collected for hypothesis testing were used for statistical analysis and for mobile business platform users. The following conclusions are drawn. (1) ESG activities have become key variables that enable sustainable corporate growth. Companies can implement eco-friendly operating processes through ESG activities. (2) This study verifies the relationship between AI-based digital transformation and corporate sustainability and confirms that digital transformation positively affects corporate sustainability. In addition, societal problems can be identified and environmental accidents prevented through technological innovation. (3) This study does not verify the positive moderating effect of green innovation; however, it emphasizes its necessity and importance. Although green innovation improves performance only in the long term, it is a key factor for companies pursuing sustainable growth. This study reveals that ESG- and AI-based digital transformation is an important tool for promoting corporate sustainability, broadening the literature in related fields and providing insights for corporate management and government policymakers to advance corporate sustainability.

# 1 INTRODUCTION

To effectively prepare for eco-friendly issues and carbon neutrality emphasized in corporate sustainability growth, active use of artificial intelligence-based digital technology is required. The core of green and digital transformation is digital new technologies such as artificial intelligence and big data, which can create an innovative, sustainable, comprehensive and resilient sustainable environment. An important interface between ESG and digital technology can be found in the environment, social, and governance. First, we need to develop new technologies based on artificial intelligence through digital transformation of eco-friendly operating processes in priority areas such as energy, mobility, agriculture and construction. Next, it is possible to secure high-quality jobs by investing in green transactions, low-emissions and digital technologies, and utilizing public and private technologies and capabilities that can be used appropriately. Finally, there is a need to establish a sustainable decision-making system, such as economic activity planning with high-quality data, data analysis models, and energy-efficient artificial intelligence-based solutions.

In the past, corporate value and sustainability were frequently assessed using quantitative metrics; however, COVID-19 and the global abnormal climate have changed the business environment for firms. Environment, social responsibility, and governance (ESG) administration issues are receiving attention. ESG is emerging as a new paradigm for management to benefit corporate sustainability, and businesses are pursuing cost and sustainable growth through voluntary and blame-based management strategies, in addition to social responsibility, to stop environmental damage. Korean businesses that lead the world market are also concentrating on ESG and integrating ethics, sustainability, and eco-friendliness into their usual business. In addition, businesses are realizing the value of ESG, expanding the range of ESG-related operations, and creating a variety of management strategies that include social responsibility, environmental, and governance structures. ESG has become a new area of business competition in response to the rapidly changing market environment. Companies create new sustainable development through ESG; however, if ESG is neglected, they may also be exposed to grave risks that endanger their existence.

ESG and digital transformation is becoming an increasingly significant strategy for managing business value and sustainability in corporate green management operations. It serves as a crucial component that can enhance creating a sustainability environment, especially in an uncertain setting such as a pandemic, as a means of preparing for the future and fostering sustainable development. It serves as a measure for assessing an organization's non-financial performance by calculating the social impact of each size category of ESG (Galbreath, 2013). However, as the exterior surroundings becomes difficult, such as Covid-19 energy shortages, and difficulties in securing supply chains, businesses are investing greater in efforts for sustainable growth than ever before. ESG and digital transformation has a positive impact on the transformation of sustainable growth in the business administration tasks.

To satisfy stakeholder needs, strengthen competitiveness, protect the environment, and achieve social ideals, businesses are adopting sustainable management, which prioritizes environmental management and transparent governance. Therefore, a corporate ESG activities are an important paradigm for corporate sustainability, and effective ESG activities boost competitiveness and enable sustainable growth. The majority of studies on ESG have focused on empirically proving the influence of ESG activities on corporate sustainability (Son and Lee, 2019; Bajic and Yurtoglu, 2018; Eccles, Ioannou and Serafeim et al., 2014; El Ghoul et al., 2018; Fatemi, Fooladi and Tehranian, 2015; Qureshi et al., 2019), corporate value (Rezaee, 2016; Tarmuji, Maelah and Tarmuji, 2016; Yu, Guo and Luu, 2018; Kang and Jung, 2020) assessment, or the impact of ESG on financial performance (Friede, Busch and Bassen, 2015; Velte, 2017; Zhao et al., 2018). However, it is crucial to comprehend ESG from the perspective of the consumer and discuss how these impacts effect the sustainability of businesses.

Consumers create a precise sustainability for businesses that actively engage in ESG activities, but for businesses that are merely playing, it may result in boycotts and resistance. This has a direct bearing on a company's revenues and may have an impact on its future growth. Hence, a company's ESG activities can be utilized as a clue to create a corporate sustainability, which has an impact on how customers perceive the company's goods and services. Therefore, it is predictable that this will have an impact on a corporate sustainable growth. Given the interdependence of social responsibility, environmental, and economic issues in a corporate ESG activities, a danger exists if any one component is ignored, making sustainable growth impossible (Holden et al., 2017). Companies should strive to enhance their corporate social responsibility activities while pursuing sustainable economic growth through ESG activities. In particular, it is crucial to manage changes in stakeholders and actively respond to areas such as abnormal climate, carbon neutrality, and supply chain management, which are exceedingly socially relevant.

This study examines how corporate sustainability is related to ESG and artificial intelligence-based digital transformation. By establishing a corporate sustainability of ESG efforts and artificial intelligence-based digital transformation, this study also intends to confirm the moderating effect of green innovation on the process of increasing sustainability. By outlining the purpose and significance of corporate ESG activities and offering an effective strategic direction for ESG and by establishing a corporate sustainability of ESG efforts and artificial intelligence-based digital transformation to create a positive corporate sustainable development, this study aims to offer fundamental information for related research. Additionally, corporate ESG and artificial intelligence-based digital transformation will provide implications as a sustainable management strategy with a positive impact on all stakeholders.

## 2 LITERATURE REVIEW AND HYPOTHESES

### 2.1 ESG

ESG refers to management activities in which businesses actively resolve and promote social issues, environmental protection, full compliance with pan-ethics, and transparent governance structure management (Wu, Xiong and Gao, 2022; Gillan, Koch and Starks, 2021; Akbari et al., 2021). ESG includes solving social problems and realizing social values facing crises, such as abnormal climate, environmental conservation, carbon neutrality, and the gap between rich and poor (Alsayegh, Abdul Rahman and Homayoun, 2020). In addition, economic aspects should be considered to build a fair society while protecting the environment (Di Simone, Petracci and Piva, 2022). ESG is a non-financial evaluation index and a management strategy that addresses issues such as environmental pr otection beyond corporate social responsibility and shared value generation (Achim and Borlea, 2015 ). ESG is seen as a crucial value that is closely related to the sustainable survival and growth do businesses. Companies can accomplish sustainable growth through ESG activities by limiting the negative effects of abnormal climate and environmental issues, achieving social objective, and maximizing the usefulness of governance (Gatti, Caruana and Snehota, 2012). In particular, as talks regarding ESG activities become more prevalent in the Corona-19 environment become more prevalent in the social value of ESG as well as the sustainable development of profits and enhancement of the corporate sustainability (Lins, Servaes and Tamayo, 2017).

ESG can provide competitive advantage by balancing social responsibility, environmental responsibility, and ethical management practices. Companies can efficiently use resources through ESG and maintain competitiveness in areas such as organizational processes, member management, and innovation management (Alsayegh, Abdul Rahman and Homayoun, 2020). Consequently, the company's competitiveness and corporate sustainability improve. Furthermore, enhancement of corporate sustainability can be evaluated as a competitive factor that generates profits (Giese et al., 2019). Moreover, the environmental, social responsibility, and governance factors included in ESG are considered essential for sustainable growth (Di Simone, Petracci and Piva, 2022).

Environmental protection, carbon neutrality, and effective energy usage are becoming crucial global challenges, in addition to economic development (Son and Lee, 2019). Therefore, nations worldwide are undertaking various measures to reduce the adverse effects of environmental issues on the growth of their respective economies. Abnormal climate and carbon emissions are two of the most significant environmental problems. Companies can boldly cut carbon emissions and advance carbon zeroing for the sustainability and survival of humanity (Dimson, Karakaş and Li, 2015). Companies should be responsible for preserving the environment, and for sustainable development, the level of environmental protection should be increased through resource and waste management to reduce environmental pollution and use energy effectively (Qureshi et al., 2019).

In terms of governance structure, it is crucial to have an open organizational structure and an extremely trustworthy audit system to realize environmental and social goals. By protecting the interests of all parties involved, a sound and transparent governance structure can minimize the incidence of extra expenses, such as inspections (Fama and Jensen, 1983; Claessens, 2006; Karwowski and Raulinajtys-Grzybek, 2021). While creating an ESG committee within the Board of Directors to handle ESG operations, companies must define the roles of the Board of Directors and Audit Committees (Shrivastava and Addas, 2014). In particular, stronger governance at higher levels is required to thwart corruption that violates business ethics and legal regulations. Companies should have a properly focused equity structure and an independent and transparent board of directors, which is externally supported by stakeholders and has a positive impact on managerial activities (Conyon and Peck, 1998; Core, Holthausen and Larcker, 1999; Yermack, 2017). Therefore, effective and transparent governance can enhance businesses' capacity to manage internal and external risks, prevent needless losses, and boost long-term growth (Ho, 2005).

**2.2 Digital Transformation**

Digital transformation is a broad sense that applies IT technology along the ethos of the Fourth Industrial Revolution and which leads to digital innovation of artificial intelligence throughout different strata of society such as at the country, institutional, and commercial levels (Llopis-Albert, Rubio and Valero, 2021). It is defined as a narrow sense, meaning it is a process of developing a new business model using digital capabilities such as big data, IoT, cloud service, artificial intelligence (AI), or taking the innovation of existing products and services to disrupt the management environment (Priyono, Moin and Putri, 2020). In order to ensure competitive advantage such as through the creation of new value and profit increase, businesses have secured competitiveness for sustainability by digitalizing existing business methods, products, and services using advanced technology (Schwertner, 2017). In addition, it means that the new technology that spearheads the Fourth Industrial Revolution is introduced to change an enterprise's operation and production methods and leads to innovation in the overall operation (Klein, 2020).

Enterprises can expect a sustainable growth that enhances their core value through big data analysis centered on consumer demand through digital transformation in addition, real-time communication between enterprises and consumers is possible, thereby creating new value and providing customized products and services to improve profits (Heavin and Power, 2018). The change in business operation methods and purchase methods through being hyper-connected can ultimately increase the efficiency of an enterprise and improve its sustainable through performance, and can cause a change in the operation environment and the creation of new added value (Vial, 2019; Wessel et al., 2021). In a rapidly changing commercial landscape and with the rise of the digital economy, it is more important than ever for enterprises to introduce new technologies to be sustainable and to strengthen competitiveness (Gil-Gomez et al., 2020).

Digital transformation extends beyond just a change of specific enterprise operation through technological innovation—it encompasses the overall change of society and the economic, social, and environmental influences caused by it (Abad-Segura et al., 2020). Enterprises are emphasizing the importance of social responsibility activities and environmental responsibility activities while pursuing future-oriented development through sustainability management (Schuler et al., 2017). In addition, through technological innovation such as artificial intelligence-based digital transformation, societal problems can be identified in advance, and new opportunities can be created to contribute to the stability and development of the community, and environmental accidents can be prevented in advance, thereby improving the sustainable through performance of environmental conservation efforts (Fernando, Jabbour and Wah, 2019). By seeking new ways of working with digital transformation that can strengthen the competitiveness of the nation and the enterprise. Digital transformation will have a more positive effect on sustainable growth that pursues social prosperity and environmental conservation beyond the economic value of the enterprise (Hysa et al., 2020).

## 2.3 Corporate Sustainability

To realize social values while considering environmental variables, including economic profit-seeking, environmental protection, the creation of eco-friendly products, and abnormal climate, sustainability should be accompanied by ethical management and transparent management (Baumgartner and Ebner, 2010). Economic profitability, environmental soundness, and social responsibility are the three pillars of sustainability, or the triple bottom line (TBL), which emphasizes that all three should be considered for long-term growth and development rather than short-term economic profit generation (Elkington, 1997). Businesses must achieve economic profitability while fulfilling their social and environmental obligations to promote sustainable growth (Erol et al., 2009). To achieve sustainable growth, companies should continuously improve their value through management activities that consider both financial performance, such as sales and profits, as well as non-financial performance, such as ethical, environmental, and social problems (Orazalin and Mahmood, 2021). Recently, as the number of consumers participating in good consumption has increased, companies are in a situation where sustainability for economic, social, and environmental responsibilities is not an option, but a necessity (Alsayegh, Abdul Rahman and Homayoun, 2020).

Corporate sustainability can be a long-term objective and is influenced by social responsibility, environmental consciousness, and behavior, in addition to economic success (Ciasullo and Troisi, 2013). Sustainability is a business activity that achieves social goals, along with the creation of economic profits obtained by companies in management activities, and pursues sustainable growth of companies through nature conservation in environmental aspects (Uyar et al., 2020). Sustainability refers to a company's management strategies in the economy, society, and the environment (Ruggerio, 2021). A sustainable corporate environment is created by balancing environmental regulations with economic growth (Hermundsdottir and Aspelund, 2021). Various management activities to minimize the efforts and risk factors of companies aiming for sustainability in the economic, social, and environmental fields and to expand corporate value improve the competitiveness of companies (Mohsin et al., 2021). Therefore, achieving sustainability entails balancing social, environmental, and economic growth (Sierra and Suárez-Collado, 2021).

## 2.4 Green Innovation

The creation of innovative new procedures and systems to stop or lessen environmental pollution is referred to as "green innovation" (Schiederig, Tietze and Herstatt, 2012). For businesses to develop a sustainable competitive edge, green innovation is necessary (Chen, Lai and Wen, 2006). Environmental preservation, energy saving, waste management, resource recovery, and the adoption of green processes are all examples of "green innovation" (Takalo and Tooranloo, 2021). Green innovation is a strategy that emphasizes resource use and pollution prevention throughout the life cycle, including products and production processes (Song and Yu, 2018). Green innovation is the process of innovating the entire value chain, developing a future-oriented process, and creating competitiveness to solve and prevent the negative image caused by the excessive waste of resources and environmental pollution (Li et al., 2018; Oltra and Saint Jean, 2009). The positive economic value of green innovation can be applied to sustainability in the future (Amores-Salvadó, Martín-de Castro and Navas-López, 2014). Green innovation is crucial for achieving outstanding productivity, and is an appropriate strategy for sustainable growth (Akao and Managi, 2007; Tolliver et al., 2021).

As the efficient utilization of resources and eco-friendly management activities can improve the reputation and corporate sustainability, green innovation is directly related to the economic profit creation and sustainable growth of the company (Farza et al., 2021). Green innovation is a means of eliminating or minimizing negative environmental impacts (Fernando, Jabbbour and Wah, 2019). Companies attempt to safeguard the environment by minimizing resource consumption and regulating waste and pollution through green innovation to boost their corporate sustainability (Rossiter and Smith, 2018). Green innovation helps companies provide eco-friendly images to consumers to reduce environmental impact (Albort-Morant, Leal-Millán and Cepeda-Carrión, 2016). As consumers' perception of environmental protection is strengthened, their demand for eco-friendly products is increasing (Song and Yu, 2018), and companies are increasing their investment in green innovation (Takalo and Tooranloo, 2021).

Companies may respond to customer demand promptly through green innovation and can develop fresh and inventive ideas through diverse engagements with consumers (Roome and Wijen, 2006). Consumers, on the other hand, are more inclined to shun firms that are unfriendly to the environment (Oltra and Saint Jean, 2009). Green innovation increases environmental consciousness, resulting in long-term profitability and eco-friendly competitiveness. Green innovation assists businesses in increasing sales via resource efficiency and the creation of new clients who are ready to pay for environmentally friendly products and services, thereby assisting a company's long-term income and development (Ahmeda, Mozammelb and Zamanc, 2020; Nadeem et al., 2020).

## 2.5 Relationship between ESG and Corporate Sustainability

According to Brown and Dacin's (1997) research, firms may create trust with local communities by consistently participating in certain social responsibility activities, further strengthen corporate sustainability, and form favorable sentiments about the companies' products. Furthermore, if they are regarded to be authentic and have conducted management operations for social issue resolution and environmental preservation, customers will build a competent and warm picture of the firm (Tarmuji, Maelah and Tarmuji, 2016). When it is decidrd that a corporation seeks both social value and economic benefit, customers favourably perceive the corporate sustainability (Koh, Burnasheva and Suh, 2022). Companies may assist in forming a competitive and efficient corporate sustainability for consumers by equitable and maintaining transparent relationships with stakeholders (Chen et al., 2018). Systematic governance provides clues that a company provides high-quality products, and consumers infer the ability and expertise of the company through these clues (Hossain, Alamgir and Alam, 2016).

Companies that engage in ESG, such as resource efficiency and environmental preservation, have a favorable sustainability because they participate in community development and achieve social value (Ramesh et al., 2019; Gürlek, Düzgün and Uygur, 2017). Furthermore, prior research has clearly proven a direct association between social responsibility activities and corporate sustainability (He and Lai, 2014; Chen et al., 2021). Han's (2021) research revealed that environmental responsibility actions might lead to the construction of positive corporate sustainability. Companies that are interested in environmental protection and abnormal climate and implement eco-friendly management activities improve their corporate sustainability (Sen, Bhattacharya and Korschun, 2006), and companies that support social responsibility activities such as social problem-solving, including job creation for vulnerable groups, improve their corporate sustainability. Furthermore, management efforts that comply with regulations and aim to avoid corruption improve the corporate sustainability (He and Lai, 2014).

This study established the following hypotheses based on the findings of prior investigations.

*Hypothesis 1: ESG activities have a positive effect on corporate sustainability.*

*Hypothesis 1-1: Environment activities have a positive effect on corporate sustainability.*
*Hypothesis 1-2: Social responsibility have a positive effect on corporate sustainability.*
*Hypothesis 1-3: Governance have a positive effect on corporate sustainability.*

## 2.6 Relationship between Digital Transformation and Corporate Sustainability

Artificial intelligence-based digital transformation is becoming an essential element for companies to attain competitiveness in the rapidly changing era of the Fourth Industrial Revolution. Businesses are trying to them improve sustainable through performance by introducing new technologies and bringing in high-tech concepts such as cloud computing, artificial intelligence, big data, and IoT to strengthen their competitive advantage. Conglomerates are creating the greatest profits by efficiently using organizational resources through digital transformation Nwankpa and Roumani (2016) verified the moderating effect of digital transformation in the relationship between IT capability and organizational performance of a U.S. enterprise, targeting the chief information officer, and discovered the influence of digital transformation on corporate sustainability is greater than the effect of innovation on corporate sustainability.

Digital transformation not only creates changes in all areas of our society; its influence is expanding, and the interest from executives is increasing. Digital transformation has contributed to environmental responsibility activities such as in the reduction of environmental waste emission and environmental preservation through big data and artificial intelligence, and as a result, it is leading to environmental sustainability (Priyono, Moin and Putri, 2020). Digital transformation fundamentally changes existing business models and operating methods, and builds a system for social problem solving, thereby increasing profits and securing competitiveness (Heavin and Power, 2018). Enterprises will expand new business areas and create new jobs, which will contribute to the employment of vulnerable groups in the community. Therefore, digital transformation is expected to play a positive effect on strengthening the corporate sustainability.

*Hypothesis 2: Digital transformation has a positive effect on corporate sustainability.*

## 2.7 Moderating Effect of Green innovation

Green innovation attempts to provide eco-friendly products and services to consumers while recognizing the importance of environmental degradation (Li et al., 2018). Green innovation has become increasingly essential, as it interacts with firms' value-generating operations. Furthermore, as governments in each country tighten environmental regulations in response to abnormal climate and consumers' awareness of the importance of using eco-friendly products, companies have strongly influenced their sustainable development by increasing their competitiveness through green innovation (Bossle et al., 2016). Green innovation is a strategy for reducing or eliminating negative environmental repercussions (Fernando, Jabbour and Wah, 2019).

Green innovation may increase a firm's value and enable sustainable development by serving as a vital competitive force in long-term development and making optimal use of limited resources (Rossiter and Smith, 2018). Green innovation helps businesses cut raw material and waste disposal expenses by utilizing resources more efficiently and recycling trash (Takalo and Tooranloo, 2021). Companies may gain a competitive edge by offering unique products and services through product innovation and the creation of new systems (Aguilera-Caracuel and Ortiz-de-Mandojana, 2013). Green innovation may reduce environmental expenses by lowering waste emissions in accordance with environmental regulations (Weng, Chen and Chen, 2015).

Finally, green innovation may increase revenue by generating eco-friendly, raising societal awareness, and setting eco-friendly premium pricing (Nadeem et al., 2020). Hence, companies may use green innovation to accomplish product innovation and process improvement, while lowering operational expenses. Particularly, as the influence of environmental problems on businesses has grown, the necessity for green innovation has become more critical than ever (Muisyo and Qin, 2021). Companies that prioritize green innovation in their management operations prioritize environmental protection, seek new solutions to environmental challenges, and invest in technological innovation to fulfill environmental goals (Zhang and Ma, 2021). Consequently, green innovation is predicted to have a favorable influence on corporate sustainability. Thus, the hypothesis for this investigation is as follows:

*Hypothesis 3: Green innovation positively moderates the relationship between the digital transformation and corporate sustainability.*

*Hypothesis 4: Green innovation positively moderates the relationship between the ESG activities and corporate sustainability.*

*Hypothesis 4-1: Green innovation positively moderates the relationship between the Environment activities and corporate sustainability.*
*Hypothesis 4-2: Green innovation positively moderates the relationship between the Social responsibility and corporate sustainability.*
*Hypothesis 4-3: Green innovation positively moderates the relationship between the Governance and corporate sustainability.*

## 3 METHODS

### 3.1 Research model and Statistical analysis

The purpose of this study was to confirm the relationship between corporate ESG, artificial intelligence-based digital transformation, and corporate sustainability. In addition, the moderating effect of green innovation were confirmed. Figure 1 illustrates our research model. This research examined the statistical programs SPSS 26.0 and AMOS 22.0. Frequency analysis was used for demographic analysis, SPSS was used for exploratory factor analysis, reliability, correlation, and hypothesis verification of measuring instruments, and AMOS was used for confirmatory factor analysis (CFA).

<Insert Figure 1 about here>

**3.2 Sample and data collection**

To achieve the purpose of this study, users using mobile business platforms (ssg.com, Lotte-on, Gmarket, Interpark, Auction, 11th Street, CJ On Style, etc.) were selected for the questionnaire survey, and an online survey was conducted. In selecting the questionnaire survey, users in their 20s and 40s who regularly use the mobile business platform were selected, and the survey was answered using self-report, which allowed the selected respondents to write directly. A total of 359 copies of the data collected for hypothesis testing were used for statistical analysis. The demographic characteristics of the samples used in this study are as follows: There were 247 women (68.8 %) and 112 men (31.2 %). The age was in the order of 30s 173 (48.2%), 20s 95(26.5%), and 40s 91 (25.3%). The academic background was high school in 72 (20.1%), graduate school in 203 (56.5%), and junior college in 84 (23.4 %). In terms of income, 73 (20.3%) earned less than 30 million won, 191 (53.2%) earned 30 million won to 50 million won, 69 (19.2%) 50 million won to 70 million won, and 26 (7.3%) more than 70 million won. There were 114 (31.8%) employees, 84 (23.4%) self-employed workers, 64 (17.8%) professionals (doctors and lawyers), 66 (18.4%) civil servants, and 31 (8.6%) others. In terms of the frequency of mobile business platform use, 121 (33.7%) used the platform 1-2 times a week, 189 (52.6%) used it 3-5 times a week, and 49 (13.7%) used it more than 5 times a week.

**3.3 Measures**

This study created assessment items by splitting ESG activities into environment, social responsibility, and governance activities to assess consumers' perceptions of the ESG activities of mobile business platform businesses (Wu, Xiong and Gao, 2022; Gillan, Koch and Starks, 2021; Akbari et al., 2021). Environment activities were defined as those associated with firms' attempts to safeguard the environment, such as resource consumption, waste emission reduction, resource conservation, eco-friendly manufacturing operations, and energy efficient usage (Alsayegh, Abdul Rahman and Homayoun, 2020; Akbari et al., 2021). The measuring tools described by Akbari et al. (2021) were used to assess environment activities.

Social responsibility activities were defined as the level of efforts to provide various activities and quality employment environments for community problem solving, protect consumer rights, and promote community win-win cooperation (Gatti, Caruana and Snehota, 2012; Akbari et al., 2021). Specifically, the measurement tool presented in Akbari et al. (2021) was used to measure social responsibility activities and consisted of five questions.

Governance activities are defined as corporate efforts such as corporate management responsibility, shareholder rights protection, or the establishment of a monitoring system for CEOs (Afzali and Kim, 2021). The measurement tool presented by Afzali and Kim (2021) was used to measure governance activity.

Digital transformation was defined as "the form of changing the operation and production method of the enterprise and causing overall organizational innovation by introducing new technology based on artificial intelligence that leads the development flow of the Fourth Industrial Revolution." The measurement tools presented in Abad-Segura et al. (2020) research were measured into four items.

Sustainability is defined as the business activities of companies aimed at sustainability in the economic, social responsibility, and environmental fields, to minimize the risk of these activities, and to promote corporate value (Erol et al., 2009; Ruggerio, 2021). To measure sustainability, the measurement tool presented in Ruggerio (2021) was used, and was composed of six questions.

Green innovation is defined as an effort to create a competitive advantage by combining ecological environment ideology with the strategic goals of companies for companies to develop new products, services, systems, and markets (Nadeem et al., 2020). Specifically, the measurement tool presented by Nadeem et al. (2020) was used to measure green innovation and was composed of five questions.

All measurement items in this study used a 7-point Likert scale (1 = strongly disagree, 7 = strongly agree).

## 4 RESULTS

### 4.1 Validity and Reliability Analysis

The construct validity of the measuring instrument was tested using convergent and discriminant validity. In addition, internal consistency reliability was assessed using Cronbach's alpha to corroborate the assessment tool's dependability. Confirmatory factor analysis (CFA) was performed using AMOS 22.0 to validate construct validity, and reliability analysis was performed using SPSS 26.0 to check reliability. Table 1 shows the results of CFA. the results were as follows: $X^2(p)$=1091.151(.000), $X^2$/df=2.864, RMSEA=.072, IFI=.948, CFI=.940, PNFI=.807, and PGFI=.668. Based on this analysis, we concluded that the research model is a suitable fit. Table 1 presents the findings of the research model's fit index.

We also examined the average variance extraction (AVE) of each concept investigated in this study to determine the convergent validity of the constructs. The results showed that the standardized regression weights of the environment ranged from .820 to .931, social responsibility from .794 to .918, governance from .646 to .974, digital transformation from .624 to .963, green innovation from .839 to .917, and sustainability from .589 to .941. In addition, the average variance extracted (AVE) for the environment was .795, social responsibility was .731, governance was .634, digital transformation was .641, green innovation was .745, and sustainability was .636. Therefore, the AVE of the variables was more than .5, and the measurement had considerable validity (Jin and Hahm, 2021). The value of composite reliability (CR) of environment was .927, social responsibility was .892, governance was .782, digital transformation was .753, green innovation was .910, and sustainability was .851. Therefore, the CR of the variables was greater than .7, and the measurement had considerable validity. Table 1 summarizes the convergent validity of the results.

Because responses were gathered from the respondents in the same manner to measure the variables in this study, the common method variance (CMV) problem might possibly arise (Podsakoff et al., 2003). Even if single or several factors are derived from exploratory factor analysis (EFA), CMV exists if it explains more than 50% of the variation in the measured variables (Fuller et al., 2016). To investigate CMV, we performed exploratory factor analysis, which allowed us to validate the eigenvalues and total variance values. Environment accounted for 16.829% of the variance with an eigenvalue of 5.049, social responsibility accounted for 15.895% of the variance with an eigenvalue of 4.769, governance accounted for 15.176% of the variance with an eigenvalue of 4.553, digital transformation accounted for 13.530% of the variance with an eigenvalue of 4.059, green innovation accounted for 12.398% of the variance with an eigenvalue of 3.719, and sustainability accounted for 10.302% of the variance with an eigenvalue of 3.091. According to these results, every eigenvalue was once greater than one, and the ordinary variance value was below 50%. Our results support the notion that CMV was not a concern in this study.

The dependability of the measuring tool was evaluated using a reliability analysis. The Cronbach's alpha was used to evaluate the reliability of the variables. The reliability analysis findings are presented as follows: Cronbach's alpha for environment was .981, social responsibility was .961, governance was .935, digital transformation was .889, green innovation wan .968, and sustainability was .944. Cronbach's alpha coefficients were all greater than .8. Nunnally (1978) proposed that dependability is significant when it higher than .7. Consequently, the dependability of the variables was considerable and genuine.

<center>**<Insert Table 1 about here>**</center>

### 4.2 Correlation Analysis

Discriminant validity is confirmed when the AVE value of each variable is higher than all correlation coefficients of values. According to the results of discriminant validity, AVE values were ENW= .795, SOC=.731, GOV=.634, EI=.641, GI=.745, and SUS=.636, which are higher than all correlation coefficients. Therefore, discriminant validity was confirmed. Table 2 presents the results of the correlation analysis.

<Insert Table 2 about here>

### 4.3 Hypothesis Test

In this study, we established eight hypotheses. SPSS (version 26.0) was used for the analysis. First, we examine the effects of ESG activities on corporate sustainability. Second, we examined the effects of digital transformation on corporate sustainability. Third, we examined how green innovation moderates the relationship between digital transformation and corporate sustainability.

Hypothesis 1-1 establishes that the environment of ESG activities positively affects corporate sustainability. Environmental management activities had a significant positive effect on corporate sustainability ($\beta$=.331, $p$<.001). Consequently, Hypothesis 1-1 is accepted, and this result explains why environmental management activities improve the corporate sustainability. Hypothesis 1-2 posits that social responsibility of ESG activities positively affect corporate sustainability. Social responsibility activities have a significant positive effect on corporate sustainability (β=.145, p<.01). Consequently, Hypothesis 1-2 is accepted, and this result explains why social responsibility activities improve the corporate sustainability. Hypothesis 1-3 established that the governance of ESG activities positively affects corporate sustainability. Governance have a significant positive effect on corporate sustainability (β=.224, p<.001). Consequently, Hypothesis 1-3 is accepted, and this result explains why governance activities improve the corporate sustainability. Table 3 shows the results for Hypothesis 1-3.

<Insert Table 3 about here>

Hypothesis 2 posits that digital transformation positively affects corporate sustainability. Digital transformation has a significant positive effect on corporate sustainability ($\beta$ =.232, $p$<.001). Consequently, Hypothesis 2 is accepted, which explains why digital transformation improves corporate sustainability. Table 4 presents the results for hypothesis 2.

<Insert Table 4 about here>

We examine the moderating effect of green innovation on the relationship between digital transformation and corporate sustainability. Regression evaluation was performed using SPSS 26.0 to confirm the hypothesis. Hypothesis 3 confirms that green innovation positively moderates the impact of digital transformation on corporate sustainability. However, the results show that green innovation negatively moderates the impact of digital transformation on corporate sustainability ($\beta$=-.293, $p$<.001). Therefore, Hypothesis 3 is rejected. The results show that the interplay between the digital transformation and green innovation decreases corporate sustainability. Table 5 presents the results for the moderating influence of green innovation.

<Insert Table 5 about here>

We examine the moderating effect of green innovation on the relationship between ESG and corporate sustainability. Regression evaluation was performed using SPSS 26.0 to confirm the hypothesis. Hypothesis 4-1 confirms that green innovation positively moderates the impact of environmental activities on corporate sustainability. However, the results show that green innovation negatively moderates the impact of environmental activities on corporate sustainability ($β=-.264$, $p<.001$). Therefore, Hypothesis 4-1 is rejected. Hypothesis 4-2 confirms that green innovation positively moderates the impact of social responsibility on corporate sustainability. However, the results show that green innovation negatively moderates the impact of social responsibility on corporate sustainability ($β=-.173$, $p<.05$). Therefore, Hypothesis 4-2 is rejected. Hypothesis 4-3 confirms that green innovation positively moderates the impact of governance on corporate sustainability. However, the results show that green innovation negatively moderates the impact of governance on corporate sustainability ($β=-.100$, $p>.05$). Therefore, Hypothesis 4-3 is rejected. The results show that the interplay between the ESG and green innovation decreases corporate sustainability. Table 6-8 presents the results for hypothesis 4.

<center><strong><Insert Table 6 about here></strong></center>
<center><strong><Insert Table 7 about here></strong></center>
<center><strong><Insert Table 8 about here></strong></center>

These results are intriguing: green innovation may lose its innovation because it requires considerable investment or time from the company's perspective. Green innovation is a lengthy process; innovation performance must be accomplished through regular investment, and innovation mechanisms must be supplemented to suit the government's policy. Therefore, it is natural for corporations to carry out green innovation; however, companies want to complement innovative activities in accordance with their characteristics and circumstances. Because green innovation is about innovating the entire value chain and overall management, the user's approach and the company's approach are different. In other words, users only use technology or products, but companies approach products or services differently, keeping in mind the entire process from research to sales and after-sales service, so continuous investment is necessary.

## 5 DISCUSSION AND CONCLUSIONS

This study validates the relationship between ESG activities, digital transformation and corporate sustainability and investigates the moderating effect of green innovation. We focused on the ESG activities of mobile business platform enterprises, because they should be particularly conscious of the significance of sustainability. We investigate the moderating role of green innovation on corporate sustainability to increase the effect of digital transformation on corporate sustainability. However, we do not demonstrate that green innovation has a moderating effect. We consolidate the study's findings linking corporate ESG, digital transformation, green innovation, and corporate sustainability based on relatively few studies that have examined ESG activities in determining corporate sustainability. We aim to identify this theory and examine the role of corporate ESG activities in corporate sustainability. Furthermore, we extend this to the field of research on ESG and simultaneously measure moderating effects. Based on these results, we contribute theoretical and practical implications as well as future research possibilities on corporate ESG activities and corporate sustainability.

### 5.1 Theoretical Implications

Most studies on ESG have focused on how ESG affected corporate sustainability (Son and Lee, 2019; Bajic and Yurtoglu, 2018; El Ghoul et al., 2018; Qureshi et al., 2019; Friede, Busch and Bassen, 2015); however, less research has been conducted regarding how ESG activities impact the sustainable growth of a company. In this context, the academic significance of this study is that it explores and identifies corporate sustainability by drawing a connection between consumers' perceptions and responses to corporate ESG activities. This study did not focus on the direct impact of corporate ESG activities on corporate sustainability; rather, it aimed to investigate specific key variables in the process by which corporate ESG activities induce sustainable growth.

First, corporate ESG activities enhanced corporate sustainability, and the enhanced corporate sustainability led consumers to choose the company's products and services, which resulted in an increase in sales. ESG activities have become a key variable that enables corporate sustainable growth. According to the results of the data analysis, corporate ESG activities positively affect corporate sustainability. This means that corporate sustainability can be enhanced when a company faithfully and actively implements ESG activities. A good company that consistently performs ESG activities to realize social values is likely to be perceived by stakeholders as competent and reliable (Koh, Burnasheva and Suh, 2022). Therefore, corporate ESG activities have been confirmed to play a key role in corporate sustainability formation.

Second, this study verifies the relationship between digital transformation and corporate sustainability, confirms that digital transformation positively affects corporate sustainability. Perceived positive digital transformation increases consumers' willingness to pay for artificial intelligence, boosts sales, and maximizes stakeholders' interests, which ultimately affects corporate sustainability (Park and Han, 2021). Digital transformation leads to digital innovation (Llopis-Albert, Rubio and Valero, 2021). Furthermore, it is expected to improve technological innovation based on artificial intelligence. It was emphasized that innovation can improve organizational performance (Fernando, Jabbour and Wah, 2019). Such positive roles will lead to corporate sustainability.

Third, the results show that corporate ESG activities positively affect corporate sustainability through the digital transformation. Green management and good relationships with stakeholders enhance corporate sustainability (Akbari et al., 2021), and companies can provide investors with trust through transparent and fair corporate governance to realize sustainable corporate growth (Aouadi and Marsat, 2018). Therefore, corporate ESG activities such as environmental protection and creating jobs for vulnerable groups play an important role in forming a positive corporate sustainability. Additionally, this positive digital transformation facilitates sustainable corporate growth. The analysis results verified a positive moderating effect of green innovation on the relationship between digital transformation and sustainable growth, but confirmed the negative moderating effect of green innovation. Green innovation refers to adopting a new type of green process to prevent excessive resource waste and environmental pollution (Schiederig, Tietze and Herstatt, 2012). However, this requires companies to spend more money and time, which can lead to negative effects on long-term corporate growth. A company can build a green image through green innovation in the short term. However, achieving sustainable growth through innovation performance is future-oriented and consistently requires complementing innovation mechanisms that are likely to put pressure on the company. Thus, the results confirmed no positive moderating effect on the relationship between digital transformation and corporate sustainability. It is evident that companies implement green innovation to achieve sustainable growth. However, they need more applicable options and complements when implementing innovative green management.

### 5.2 Practical Implications

In addition to the theoretical contributions described above, we provide the following practical implications based on our findings.

First, online retail companies in Korea and other countries have put effort into ESG activities by establishing new business models for their inventory. Managing the inventory produced by failed demand forecasting can result in environmental contamination, additional costs, and resource waste. Therefore, transactions related to such inventory, including products returned because customers changed their minds, products displayed in the store, and partially repaired products with small scratches from online and mobile retail platforms, contribute to promoting the repeated use of existing resources, sustainable consumption, and resource recycling. This positively affects corporate performance and plays an important role as a management strategy for increasing corporate value. With the advent of a rapidly changing market environment, corporate ESG activities have emerged as a new type of business competitiveness. A company can create new business opportunities, establish a preemptive management strategy, and increase corporate value through ESG activities. Companies began to put effort into implementing systematic and efficient ESG strategies after they recognized the importance of ESG during the COVID-19 pandemic. Such companies should establish new ESG decision-making organizations, such as ESG committees, ESG implementation organizations, and working groups to facilitate the systematic implementation of their ESG management.

Second, companies should monitor ESG-related sustainability certification and introduce it in relation to their products to obtain objective indicators of ESG levels and sustainability. Because consumers who consider corporate ESG (e.g., environmental and ethical activities) in deciding their purchases regard a green and artificial intelligence-based digital transformation as important, companies should provide various values and implement balanced ESG. As this study confirms, corporate ESG positively affect corporate sustainability. Moreover, companies can efficiently utilize resources, minimize problems caused by climate change and environmental issues, realize social values, maximize benefits from transparent governance, and ultimately promote sustainable growth.

Third, it could not verify the positive moderating effect of green innovation. However, this study emphasizes the necessity and importance of green innovation. Although green innovation produces performance only in the long term, it is a key factor for companies pursuing sustainable growth. Because the efficient use of resources and green management can enhance corporate reputation and sustainability, green innovation enables companies to generate economic profits and achieve sustainable growth (Farza et al., 2021). As more consumers recognize environmental issues, there is a higher demand for green products (Song and Yu, 2018), which leads companies to increase investment in green innovation (Takalo and Tooranloo, 2021). In summary, green innovation contributes to corporate development in a sustainable economic environment despite its negative aspects in terms of cost and time. In addition, the government must preemptively expand its support for green innovation.

### 5.3 Limitations and Future Research

This study validates the relationship between corporate ESG, digital transformation and corporate sustainability and investigates the moderating effect of green innovation. Despite various theoretical and practical discussions conducted in the study, there are a few limitations, as given below.

First, we examined whether various variables mediate the relationship between ESG and corporate sustainability. Corporate value, sustainable through performance, price fairness, and perceived responses mediate the relationship between ESG and corporate sustainability. Based on the above, we advocate proving these mediating variables in future studies.

Second, this study focuses on key factors such as corporate ESG, digital transformation, green innovation, and corporate sustainability. However, it is not possible to verify the moderating effect of green innovation on the relationship between digital transformation and corporate sustainability. Thus, a discussion of various moderating variables that increase corporate sustainability through ESG is needed in future studies.

Third, the survey was administered through self-report. We contemplate that the self-reporting approach has a flaw in that the correlation between the variables is too strong. Although we completed CMV testing, we considered that common method bias (CMB) exists. Future studies are required to advance and build a data-collection method to avoid the CMB problem.

**Data Availability Statement**

The raw data supporting the conclusions of this article will be made available by the authors, without undue reservation.

**Author Contributions**

CQ performed the data collection and analysis. SJ contributed to drafting, review, and editing. Both authors contributed to the study's conception and design

**Conflict of Interest**

The authors declare that the research was conducted in the absence of any commercial or financial relationships that could be construed as a potential conflict of interest.

**Publisher's Note**

All claims expressed in this article are solely those of the authors and do not necessarily represent those of their affiliated organizations, or those of the publisher, the editors and the reviewers. Any product that may be evaluated in this article, or claim that may be made by its manufacturer, is not guaranteed or endorsed by the publisher.

**References**

Abad-Segura, E., González-Zamar, M. D., Infante-Moro, J. C., & Ruipérez García, G. (2020). Sustainable management of digital transformation in higher education: Global research trends. *Sustainability*, *12*(5), 2107. https://doi.org/10.3390/su12052107

Achim, M. V. and Borlea, S. N. (2015). Developing of ESG score to assess the non-financial performances in Romanian companies. *Procedia Economics and Finance*, *32*, 1209-1224. https://doi.org/10.1016/S2212-5671(15)01499-9

Afzali, H. and Kim, S. S. (2021). Consumers' responses to corporate social responsibility: The mediating role of CSR authenticity. *Sustainability*, 13(4), 2224. https://doi.org/10.3390/su13042224

Aguilera-Caracuel, J. and Ortiz-de-Mandojana, N. (2013). Green innovation and financial performance: An institutional approach. *Organization & Environment*, 26(4), 365-385. https://doi.org/10.1177/1086026613507931

Ahmeda, U., Mozammelb, S. and Zamanc, F. (2020). Green HRM and green innovation: Can green transformational leadership moderate: Case of pharmaceutical firms in Australia. *Systematic Reviews in Pharmacy*, 11(7), 616-617. https://dx.doi.org/10.31838/srp.2020.7.86

Akao, K. I. and Managi, S. (2007). Feasibility and optimality of sustainable growth under materials balance. *Journal of Economic Dynamics and Control*, *31*(12), 3778-3790. https://doi.org/10.1016/j.jedc.2007.01.013

Akbari, M., Omrane, A., Hoseinzadeh, A. and Nikookar, H. (2021). Effects of innovation on corporate performance of manufacturing companies: Which roles associated to social responsibility?. *Transnational Corporations Review*, 1-16. https://doi.org/10.1080/19186444.2021.1940055

Albort-Morant, G., Leal-Millán, A. and Cepeda-Carrión, G. (2016). The antecedents of green innovation performance: A model of learning and capabilities. *Journal of Business Research*, 69(11), 4912-4917. https://doi.org/10.1016/j.jbusres.2016.04.052

Alsayegh, M. F., Abdul Rahman, R. and Homayoun, S. (2020). Corporate economic, environmental, and social sustainability performance transformation through ESG disclosure. *Sustainability*, 12(9), 3910. https://doi.org/10.3390/su12093910

Amores-Salvadó, J., Martín-de Castro, G. and Navas-López, J. E. (2014). Green corporate image: Moderating the connection between environmental product innovation and firm performance. *Journal of Cleaner Production*, 83, 356-365. https://doi.org/10.1016/j.jclepro.2014.07.059

Aouadi, A. and Marsat, S. (2018). Do ESG controversies matter for firm value? Evidence from international data. *Journal of Business Ethics*, 151(4), 1027-1047. https://doi.org/10.1007/s10551-016-3213-8

Bajic, S. and Yurtoglu, B. (2018). Which aspects of CSR predict firm market value?. *Journal of Capital Markets Studies*, 2(1), 50-69. https://doi.org/10.1108/JCMS-10-2017-0002

Baumgartner, R. J. and Ebner, D. (2010). Corporate sustainability strategies: Sustainability profiles and maturity levels. *Sustainable development*, 18(2), 76-89. https://doi.org/10.1002/sd.447

Bossle, M. B., de Barcellos, M. D., Vieira, L. M. and Sauvée, L. (2016). The drivers for adoption of eco-innovation. *Journal of Cleaner production*, 113, 861-872. https://doi.org/10.1016/j.jclepro.2015.11.033

Brown, T. J. and Dacin, P. A. (1997). The company and the product: Corporate associations and consumer product responses. *Journal of marketing*, 61(1), 68-84. https://doi.org/10.1177/002224299706100106

Chen, C. C., Khan, A., Hongsuchon, T., Ruangkanjanases, A., Chen, Y. T., Sivarak, O. and Chen, S. C. (2021). The role of corporate social responsibility and corporate image in times of crisis: The mediating role of customer trust. *International Journal of Environmental Research and Public Health*, 18(16), 8275. https://doi.org/10.3390/ijerph18168275

Chen, X., Huang, R., Yang, Z. and Dube, L. (2018). CSR types and the moderating role of corporate competence. *European Journal of Marketing*, 52(7-8), 1358-1386. https://doi.org/10.1108/EJM-12-2016-0702

Chen, Y. S., Lai, S. B. and Wen, C. T. (2006). The influence of green innovation performance on corporate advantage in Taiwan. *Journal of business ethics*, 67(4), 331-339. https://doi.org/10.1007/s10551-006-9025-5

Ciasullo, M. V. and Troisi, O. (2013). Sustainable value creation in SMEs: A case study. *The TQM Journal*, 25(1), 44-61. https://doi.org/10.1108/17542731311286423

Claessens, S. (2006). Corporate governance and development. *The World bank research observer*, 21(1), 91-122. https://doi.org/10.1093/wbro/lkj004

Conyon, M. J. and Peck, S. I. (1998). Board control, remuneration committees, and top management compensation. *Academy of management journal*, 41(2), 146-157. https://doi.org/10.5465/257099

Core, J. E., Holthausen, R. W. and Larcker, D. F. (1999). Corporate governance, chief executive officer compensation, and firm performance. *Journal of financial economics*, 51(3), 371-406. https://doi.org/10.1016/S0304-405X(98)00058-0

Di Simone, L., Petracci, B. and Piva, M. (2022). Economic sustainability, innovation, and the ESG factors: An empirical investigation. *Sustainability*, 14(4), 2270. https://doi.org/10.3390/su14042270

Dimson, E., Karakaş, O. and Li, X. (2015). Active ownership. *The Review of Financial Studies*, 28(12), 3225-3268. https://doi.org/10.1093/rfs/hhv044

Eccles, R. G., Ioannou, I., and Serafeim, G. (2014). The impact of corporate sustainability on organizational processes and performance. *Management science*, 60(11), 2835-2857. https://doi.org/10.1287/mnsc.2014.1984

El Ghoul, S., Guedhami, O., Kim, H., and Park, K. (2018). Corporate environmental responsibility and the cost of capital: International evidence. *Journal of Business Ethics*, 149(2), 335-361. https://doi.org/10.1007/s10551-015-3005-6

Elkington, J. (1997). The triple bottom line. *Environmental management: Readings and Cases*, 2, 49-66.

Erol, I., Cakar, N., Erel, D. and Sari, R. (2009). Sustainability in the Turkish retailing industry. *Sustainable Development*, 17(1), 49-67. https://doi.org/10.1002/sd.369

Fama, E. F. and Jensen, M. C. (1983). Separation of ownership and control. *The Journal of Law and Economics*, 26(2), 301-325. https://doi.org/10.1086/467037

Farza, K., Ftiti, Z., Hlioui, Z., Louhichi, W. and Omri, A. (2021). Does it pay to go green? The environmental innovation effect on corporate financial performance. *Journal of Environmental Management*, 300, 113695. https://doi.org/10.1016/j.jenvman.2021.113695

Fatemi, A., Fooladi, I., and Tehranian, H. (2015). Valuation effects of corporate social responsibility. *Journal of Banking & Finance*, 59, 182-192. https://doi.org/10.1016/j.jbankfin.2015.04.028

Fernando, Y., Jabbour, C. J. C. and Wah, W. X. (2019). Pursuing green growth in technology firms through the connections between environmental innovation and sustainable business performance: does service capability matter?. *Resources, Conservation and Recycling*, 141, 8-20. https://doi.org/10.1016/j.resconrec.2018.09.031

Fernando, Y., Jabbour, C. J. C., & Wah, W. X. (2019). Pursuing green growth in technology firms through the connections between environmental innovation and sustainable business performance: does service capability matter?. *Resources, Conservation and Recycling*, *141*, 8-20. https://doi.org/10.1016/j.resconrec.2018.09.031

Friede, G., Busch, T., and Bassen, A. (2015). ESG and financial performance: Aggregated evidence from more than 2000 empirical studies. *Journal of Sustainable Finance & Investment*, 5(4), 210-233. https://doi.org/10.1080/20430795.2015.1118917

Fuller, C. M., Simmering, M. J., Atinc, G., Atinc, Y. and Babin, B. J. (2016). Common methods variance detection in business research. *Journal of Business Research*, 69(8), 3192-3198. https://doi.org/10.1016/j.jbusres.2015.12.008

Galbreath, J. (2013). ESG in focus: The Australian evidence. *Journal of Business Ethics*, 118(3), 529-541. https://doi.org/10.1007/s10551-012-1607-9

Gatti, L., Caruana, A. and Snehota, I. (2012). The role of corporate social responsibility, perceived quality and corporate reputation on purchase intention: Implications for brand management. *Journal of Brand Management*, 20(1), 65-76. https://doi.org/10.1057/bm.2012.2

Giese, G., Lee, L. E., Melas, D., Nagy, Z. and Nishikawa, L. (2019). Foundations of ESG investing: How ESG affects equity valuation, risk, and performance. *The Journal of Portfolio Management*, 45(5), 69-83. https://doi.org/10.3905/jpm.2019.45.5.069

Gil-Gomez, H., Guerola-Navarro, V., Oltra-Badenes, R., & Lozano-Quilis, J. A. (2020). Customer relationship management: digital transformation and sustainable business model innovation. *Economic research-Ekonomska istraživanja*, *33*(1), 2733-2750. https://doi.org/10.1080/1331677X.2019.1676283

Gillan, S. L., Koch, A. and Starks, L. T. (2021). Firms and social responsibility: A review of ESG and CSR research in corporate finance. *Journal of Corporate Finance*, 66, 101889. https://doi.org/10.1016/j.jcorpfin.2021.101889

Gürlek, M., Düzgün, E. and Uygur, S. M. (2017). How does corporate social responsibility create customer loyalty? The role of corporate image. *Social Responsibility Journal*, 13(3), 409-427. https://doi.org/10.1108/SRJ-10-2016-0177

Han, H. (2021). Consumer behavior and environmental sustainability in tourism and hospitality: A review of theories, concepts, and latest research. *Journal of Sustainable Tourism*, 29(7), 1021-1042. https://doi.org/10.1080/09669582.2021.1903019

He, Y. and Lai, K. K. (2014). The effect of corporate social responsibility on brand loyalty: The mediating role of brand image. *Total Quality Management & Business Excellence*, 25(3-4), 249-263. https://doi.org/10.1080/14783363.2012.661138

Heavin, C., & Power, D. J. (2018). Challenges for digital transformation–towards a conceptual decision support guide for managers. *Journal of Decision Systems*, *27*(sup1), 38-45. https://doi.org/10.1080/12460125.2018.1468697

Heavin, C., & Power, D. J. (2018). Challenges for digital transformation–towards a conceptual decision support guide for managers. *Journal of Decision Systems*, *27*(sup1), 38-45. https://doi.org/10.1080/12460125.2018.1468697

Hermundsdottir, F. and Aspelund, A. (2021). Sustainability innovations and firm competitiveness: A review. *Journal of Cleaner Production*, 280, 124715. https://doi.org/10.1016/j.jclepro.2020.124715

Ho, C. K. (2005). Corporate governance and corporate competitiveness: An international analysis. *Corporate Governance: An International Review*, 13(2), 211-253. https://doi.org/10.1111/j.1467-8683.2005.00419.x

Holden, E., Linnerud, K., Banister, D., Schwanitz, V. J. and Wierling, A. (2017). *The imperatives of sustainable development: needs, justice, limits*. Routledge.

Hossain, M. M., Alamgir, M. and Alam, M. (2016). The mediating role of corporate governance and corporate image on the CSR-FP link: Evidence from a developing country. *Journal of General Management*, 41(3), 33-51. https://doi.org/10.1177/030630701604100303

Hysa, E., Kruja, A., Rehman, N. U., & Laurenti, R. (2020). Circular economy innovation and environmental sustainability impact on economic growth: An integrated model for sustainable development. *Sustainability*, *12*(12), 4831. https://doi.org/10.3390/su12124831

Jin, X. and Hahm, S. (2021). Using online information support to decrease stress, anxiety, and depression. *KSII Transactions on Internet and Information Systems* (TIIS), 15(8), 2944-2958. https://doi.org/10.3837/tiis.2021.08.013

Kang, W., and Jung, M. (2020). Effect of ESG management and firm's financial characteristics. *Korean Journal of Financial Studies*, 49(5), 681-707. https://doi.org/10.26845/KJFS.2020.10.49.5.681

Karwowski, M. and Raulinajtys-Grzybek, M. (2021). The application of corporate social responsibility (CSR) actions for mitigation of environmental, social, corporate governance (ESG) and reputational risk in integrated reports. *Corporate Social Responsibility and Environmental Management*, 28(4), 1270-1284. https://doi.org/10.1002/csr.2137

Klein, M. (2020). Leadership characteristics in the era of digital transformation. *Business & Management Studied: AN International Journal*, 8(1), 883-902. https://doi.org/10.15295/bmij.v8i1.1441

Koh, H. K., Burnasheva, R. and Suh, Y. G. (2022). Perceived ESG (environmental, social, governance) and consumers' responses: The mediating role of brand credibility, brand image, and perceived quality. *Sustainability*, 14(8), 4515. https://doi.org/10.3390/su14084515

Li, D., Zhao, Y., Zhang, L., Chen, X. and Cao, C. (2018). Impact of quality management on green innovation. *Journal of Cleaner Production*, 170, 462-470. https://doi.org/10.1016/j.jclepro.2017.09.158

Lins, K. V., Servaes, H. and Tamayo, A. (2017). Social capital, trust, and firm performance: The value of corporate social responsibility during the financial crisis. T*he Journal of Finance*, 72(4), 1785-1824. https://doi.org/10.1111/jofi.12505

Llopis-Albert, C., Rubio, F., & Valero, F. (2021). Impact of digital transformation on the automotive industry. *Technological forecasting and social change*, *162*, 120343. https://doi.org/10.1016/j.techfore.2020.120343

Mills, J., Purchase, V. C. and Parry, G. (2013). Enterprise imaging: Representing complex multi-organizational service enterprises. *International Journal of Operations & Production Management*, 33(2), 159-180. https://doi.org/10.1108/01443571311295617

Mohsin, M., Zhu, Q., Naseem, S., Sarfraz, M. and Ivascu, L. (2021). Mining industry impact on environmental sustainability, economic growth, social interaction, and public health: An application of semi-quantitative mathematical approach. *Processes*, 9(6), 972. https://doi.org/10.3390/pr9060972

Muisyo, P. K. and Qin, S. (2021). Enhancing the FIRM'S green performance through green HRM: The moderating role of green innovation culture. *Journal of Cleaner Production*, 289, 125720. https://doi.org/10.1016/j.jclepro.2020.125720

Nadeem, M. A., Liu, Z., Ali, H. S., Younis, A., Bilal, M. and Xu, Y. (2020). Innovation and sustainable development: Does aid and political instability impede innovation?. *SAGE Open*, 10(4), 2158244020973021. https://doi.org/10.1177/2158244020973021

Nunnally, J. C. (1978). An overview of psychological measurement. *Clinical Diagnosis of Mental Disorders*, 97-146. https://doi.org/10.1007/978-1-4684-2490-4_4

Nwankpa, J. K., & Roumani, Y. (2016). IT capability and digital transformation: A firm performance perspective.

Oltra, V. and Saint Jean, M. (2009). Sectoral systems of environmental innovation: An application to the French automotive industry. *Technological Forecasting and Social Change*, 76(4), 567-583. https://doi.org/10.1016/j.techfore.2008.03.025

Orazalin, N. and Mahmood, M. (2021). Toward sustainable development: Board characteristics, country governance quality, and environmental performance. *Business Strategy and the Environment*, 30(8), 3569-3588. https://doi.org/10.1002/bse.2820

Park, Y. N. and Han, S. L. (2021). The Effect of ESG management on Corporate Image, Perceived Price Fairness, and Consumer Responses. *Korean Management Review*, 50(3), 643-664. https://doi.org/10.17287/kmr.2021.50.3.643

Podsakoff, P. M., MacKenzie, S. B., Lee, J. Y. and Podsakoff, N. P. (2003). Common method biases in behavioral research: A critical review of the literature and recommended remedies. *Journal of Applied Psychology*, 88(5), 879-903. https://doi.org/10.1037/0021-9010.88.5.879

Priyono, A., Moin, A., & Putri, V. N. A. O. (2020). Identifying digital transformation paths in the business model of SMEs during the COVID-19 pandemic. *Journal of Open Innovation: Technology, Market, and Complexity*, *6*(4), 104. https://doi.org/10.3390/joitmc6040104

Priyono, A., Moin, A., & Putri, V. N. A. O. (2020). Identifying digital transformation paths in the business model of SMEs during the COVID-19 pandemic. *Journal of Open Innovation: Technology, Market, and Complexity*, *6*(4), 104. https://doi.org/10.3390/joitmc6040104

Qureshi, D., Nayak, S. K., Maji, S., Anis, A., Kim, D. and Pal, K. (2019). Environment sensitive hydrogels for drug delivery applications. *European Polymer Journal*, 120, 109220. https://doi.org/10.1016/j.eurpolymj.2019.109220

Ramesh, K., Saha, R., Goswami, S. and Dahiya, R. (2019). Consumer's response to CSR activities: Mediating role of brand image and brand attitude. *Corporate Social Responsibility and Environmental Management*, 26(2), 377-387. https://doi.org/10.1002/csr.1689

Rezaee, Z. (2016). Business sustainability research: A theoretical and integrated perspective. *Journal of Accounting Literature*, 36(1), 48-64. https://doi.org/10.1016/j.acclit.2016.05.003

Roome, N. and Wijen, F. (2006). Stakeholder power and organizational learning in corporate environmental management. *Organization Studies*, 27(2), 235-263. https://doi.org/10.1177/0170840605057669

Rossiter, W. and Smith, D. J. (2018). Green innovation and the development of sustainable communities: The case of Blueprint Regeneration's Trent Basin development. *The International Journal of Entrepreneurship and Innovation*, 19(1), 21-32. https://doi.org/10.1177/1465750317751989

Ruggerio, C. A. (2021). Sustainability and sustainable development: A review of principles and definitions. *Science of the Total Environment*, 786, 147481. https://doi.org/10.1016/j.scitotenv.2021.147481

Schiederig, T., Tietze, F. and Herstatt, C. (2012). Green innovation in technology and innovation management: An exploratory literature review. *R&d Management*, 42(2), 180-192. https://doi.org/10.1111/j.1467-9310.2011.00672.x

Schuler, D., Rasche, A., Etzion, D., & Newton, L. (2017). Guest editors' introduction: Corporate sustainability management and environmental ethics. *Business Ethics Quarterly*, 27(2), 213-237. https://doi.org/10.1017/beq.2016.80

Schwertner, K. (2017). Digital transformation of business. *Trakia Journal of Sciences*, *15*(1), 388-393. https://doi.org/10.15547/tjs.2017.s.01.065

Sen, S., Bhattacharya, C. B. and Korschun, D. (2006). The role of corporate social responsibility in strengthening multiple stakeholder relationships: A field experiment. *Journal of the Academy of Marketing science*, 34(2), 158-166. https://doi.org/10.1177/0092070305284978

Shrivastava, P. and Addas, A. (2014). The impact of corporate governance on sustainability performance. *Journal of Sustainable Finance & Investment*, 4(1), 21-37. https://doi.org/10.1080/20430795.2014.887346

Sierra, J. and Suárez-Collado, Á. (2021). Understanding economic, social, and environmental sustainability challenges in the Global South. *Sustainability*, 13(13), 7201. https://doi.org/10.3390/su13137201

Son, S. H., and Lee, J. H. (2019). Price Impact of ESG Scores: Evidence from Korean Retail Firms. *Journal of Distribution Science*, 17(7), 55-63 https://doi.org/10.15722/jds.17.7.201907.55

Song, W. and Yu, H. (2018). Green innovation strategy and green innovation: The roles of green creativity and green organizational identity. *Corporate Social Responsibility and Environmental Management*, 25(2), 135-150. https://doi.org/10.1002/csr.1445

Takalo, S. K. and Tooranloo, H. S. (2021). Green innovation: A systematic literature review. *Journal of Cleaner Production*, 279, 122474. https://doi.org/10.1016/j.jclepro.2020.122474

Tarmuji, I., Maelah, R., and Tarmuji, N. H. (2016). The impact of environmental, social and governance practices (ESG) on economic performance: Evidence from ESG score. *International Journal of Trade, Economics and Finance*, 7(3), 67. https://doi.org/10.18178/ijtef.2016.7.3.501

Tolliver, C., Fujii, H., Keeley, A. R. ans Managi, S. (2021). Green innovation and finance in Asia. *Asian Economic Policy Review*, 16(1), 67-87. https://doi.org/10.1111/aepr.12320

Uyar, A., Kilic, M., Koseoglu, M. A., Kuzey, C. and Karaman, A. S. (2020). The link among board characteristics, corporate social responsibility performance, and financial performance: Evidence from the hospitality and tourism industry. *Tourism Management Perspectives*, 35, 100714. https://doi.org/10.1016/j.tmp.2020.100714

Velte, P. (2017). Does ESG performance have an impact on financial performance? Evidence from Germany. *Journal of Global Responsibility*, 8(2), 169-178. https://doi.org/10.1108/JGR-11-2016-0029

Vial, G. (2019). Understanding digital transformation: A review and a research agenda. *The journal of strategic information systems*, 28(2), 118-144. https://doi.org/10.1016/j.jsis.2019.01.003

Weng, H. H., Chen, J. S. and Chen, P. C. (2015). Effects of green innovation on environmental and corporate performance: A stakeholder perspective. *Sustainability*, 7(5), 4997-5026. https://doi.org/10.3390/su7054997

Wessel, L., Baiyere, A., Ologeanu-Taddei, R., Cha, J., & Blegind-Jensen, T. (2021). Unpacking the difference between digital transformation and IT-enabled organizational transformation. *Journal of the Association for Information Systems*, 22(1), 102-129. https://doi.org/10.17705/1jais.00655

Wu, C., Xiong, X. and Gao, Y. (2022). Does ESG Certification Improve Price Efficiency in the Chinese Stock Market?. *Asia-Pacific Financial Markets*, 29(1), 97-122.

Yermack, D. (2017). Corporate governance and blockchains. *Review of finance*, 21(1), 7-31. https://doi.org/10.1093/rof/rfw074

Yu, E. P. Y., Guo, C. Q., and Luu, B. V. (2018). Environmental, social and governance transparency and firm value. *Business Strategy and the Environment*, 27(7), 987-1004. https://doi.org/10.1002/bse.2047

Zhang, Q. and Ma, Y. (2021). The impact of environmental management on firm economic performance: The mediating effect of green innovation and the moderating effect of environmental leadership. *Journal of Cleaner Production*, 292, 126057. https://doi.org/10.1016/j.jclepro.2021.126057

Zhao, C., Guo, Y., Yuan, J., Wu, M., Li, D., Zhou, Y. and Kang, J. (2018). ESG and corporate financial performance: Empirical evidence from China's listed power generation companies. *Sustainability*, 10(8), 2607. https://doi.org/10.3390/su10082607

**Figure Legends**

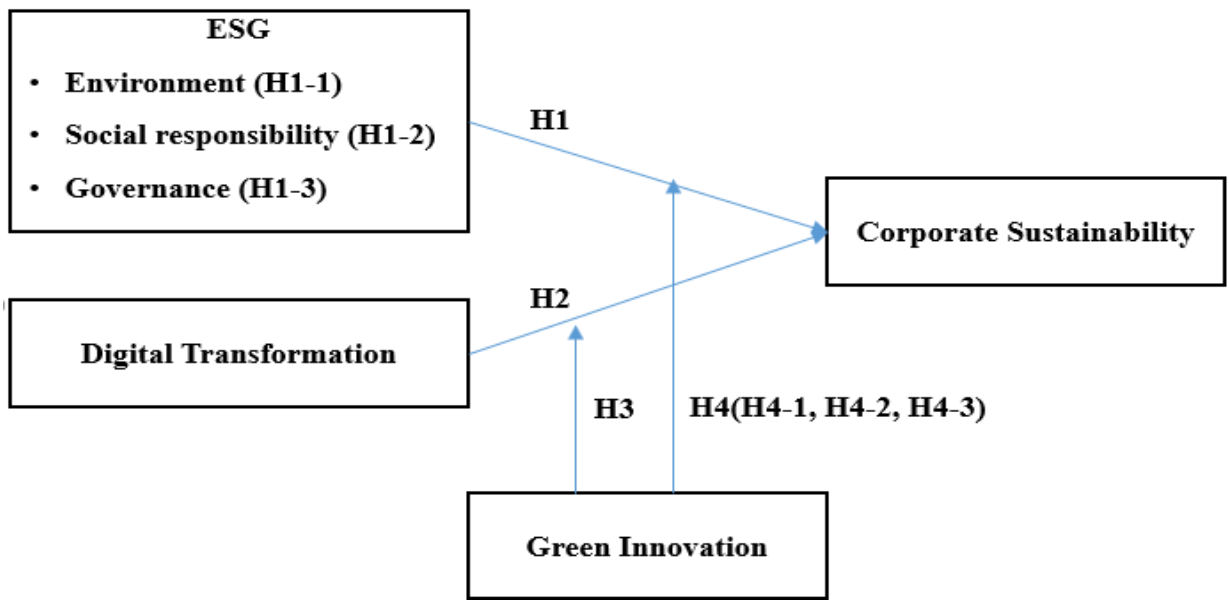

**Figure 1|** Research Model

**Tables**

**Table 1| Convergent Validity Analysis**

| Variables | Item | Estimate | S.E. | C.R. | p | Standardized Regression Weights | AVE | C.R | Cronbach's alpha |
|---|---|---|---|---|---|---|---|---|---|
| ENV | ENV1 | 1 | | | | 0.82 | .795 | .927 | .981 |
| | ENV2 | 1.128 | 0.039 | 28.893 | *** | 0.927 | | | |
| | ENV3 | 1.171 | 0.035 | 33.244 | *** | 0.909 | | | |
| | ENV4 | 1.148 | 0.039 | 29.274 | *** | 0.931 | | | |
| | ENV5 | 1.055 | 0.043 | 24.304 | *** | 0.865 | | | |
| SOC | SOC1 | 1 | | | | 0.794 | .731 | .892 | .961 |
| | SOC2 | 1.099 | 0.037 | 29.541 | *** | 0.849 | | | |
| | SOC3 | 1.144 | 0.043 | 26.849 | *** | 0.918 | | | |
| | SOC4 | 1.226 | 0.052 | 23.688 | *** | 0.868 | | | |
| | SOC5 | 0.998 | 0.045 | 22.101 | *** | 0.842 | | | |
| GOV | GOV1 | 1 | | | | 0.649 | .634 | .782 | .935 |
| | GOV2 | 0.987 | 0.051 | 19.326 | *** | 0.646 | | | |
| | GOV3 | 1.122 | 0.047 | 23.784 | *** | 0.736 | | | |
| | GOV4 | 1.483 | 0.055 | 26.977 | *** | 0.974 | | | |
| | GOV5 | 1.384 | 0.056 | 24.574 | *** | 0.916 | | | |
| DT | CI1 | 1 | | | | 0.624 | .641 | .753 | .889 |
| | CI2 | 1.025 | 0.042 | 24.381 | *** | 0.71 | | | |
| | CI3 | 1.307 | 0.053 | 24.574 | *** | 0.963 | | | |
| | CI4 | 1.25 | 0.059 | 21.023 | *** | 0.862 | | | |
| GI | GI1 | 1 | | | | 0.84 | .745 | .910 | .968 |
| | GI2 | 0.995 | 0.035 | 28.451 | *** | 0.855 | | | |
| | GI3 | 1.074 | 0.048 | 22.548 | *** | 0.839 | | | |
| | GI4 | 1.050 | 0.039 | 26.757 | *** | 0.917 | | | |
| | GI5 | 1.012 | 0.044 | 22.793 | *** | 0.861 | | | |
| SUS | SUS1 | 1 | | | | 0.824 | .636 | .851 | .944 |
| | SUS2 | 1.107 | 0.043 | 25.501 | *** | 0.832 | | | |
| | SUS3 | 1.265 | 0.043 | 29.113 | *** | 0.941 | | | |
| | SUS4 | 1.151 | 0.044 | 25.986 | *** | 0.896 | | | |
| | SUS5 | 0.877 | 0.062 | 14.079 | *** | 0.639 | | | |
| | SUS6 | 0.807 | 0.064 | 12.575 | *** | 0.589 | | | |
| Model Fit Index | $X^2$ (p)=1091.151(.000), $X^2$/df=2.864, RMSEA=.072, IFI=.948, TLI=.940, CFI=.947, PGFI=.668, PNFI=.807 | | | | | | | | |

*Note:* ***: $p < 0.001$. ENV, environment; SOC, social responsibility; GOV, governance; DT, digital transformation; GI, green innovation; SUS, corporate sustainability

**Table 2| Correlation and Discriminant Analysis**

| | ENW | SOC | GOV | DT | GI | SUS |
|---|---|---|---|---|---|---|
| ENW | **(.795)** | | | | | |
| SOC | (.198) .445 | **(.731)** | | | | |
| GOV | (.060) .244 | (.031) .177 | **(.634)** | | | |
| DT | (.110) .331 | (.021) .145 | (.050) .224 | **(.641)** | | |
| GI | (.448) .669 | (.408) .639 | (091) .301 | (.072) .269 | **(.745)** | |
| SUS | (.061) .247 | (.089) .299 | (.075) .273 | (.054) .232 | (.070) .264 | **(.636)** |

*Note:* The diagonal elements in bold represent the square root of AVE. ENV, environment; SOC, social responsibility; GOV, governance; DT, digital transformation; GI, green innovation; SUS, corporate sustainability.

**Table 3|** The influence of ESG on corporate sustainability

|  | B | Std. Error | $\beta$ | $t$ | $p$ | VIF |
|---|---|---|---|---|---|---|
| (Constant) | 3.376 | .309 |  | 10.912 | .000 |  |
| EVN | .245 | .037 | .331 | 6.633 | .000 | 1.292 |
| SOC | .126 | .045 | .145 | 2.778 | .006 | 1.255 |
| GOV | .198 | .045 | .224 | 4.345 | .000 | 1.070 |
| $F=17.963$ ($p<.001$), $R^2=.132$, Adjusted $R^2 = .124$ ||||||||

*Note:* Dependent variable: SUS. ENV: environment; SOC: social responsibility; GOV: governance; SUS: corporate sustainability.

**Table 4|** The influence of digital transformation on corporate sustainability

|  | B | Std. Error | $\beta$ | $t$ | $p$ | VIF |
|---|---|---|---|---|---|---|
| (Constant) | 3.376 | .309 |  | 10.912 | .000 |  |
| DT | .251 | .056 | .232 | 4.506 | .000 | 1.000 |
| $F=20.307$ ($p<.001$), $R^2=.054$, Adjusted $R^2 = .051$ ||||||||

*Note:* Dependent variable: SUS. DT, digital transformation; SUS, corporate sustainability

**Table 5|** The Moderating effect of green innovation (DT)

|  | Model 1 ||| Model 2 ||| Model 3 ||| |
|---|---|---|---|---|---|---|---|---|---|---|
|  | $\beta$ | $t$ | $p$ | $\beta$ | $t$ | $p$ | $\beta$ | $t$ | $p$ | VIF |
| DT (A) | 0.232 | 4.506 | .000 | 0.174 | 3.324 | .001 | 0.133 | 2.619 | .009 | 1.100 |
| GI (B) |  |  |  | 0.217 | 4.148 | .000 | 0.135 | 2.597 | .010 | 1.166 |
| Interaction (A × B) |  |  |  |  |  |  | -0.293 | -5.695 | .000 | 1.134 |
| $R^2$ (Adjusted $R^2$) | 0.054 (0.051) ||| 0.097 (0.092) ||| 0.173 (0.166) ||| |
| $F$ | 20.307 ($p<.001$) ||| 19.219 ($p<.001$) ||| 24.756 ($p<.001$) ||| |

*Note:* Dependent variable: SUS. DT, digital transformation; GI, green innovation; SUS, corporate sustainability

**Table 6|** The Moderating effect of green innovation (ENV)

|  | Model 1 ||| Model 2 ||| Model 3 ||| |
|---|---|---|---|---|---|---|---|---|---|---|
|  | $\beta$ | $t$ | $p$ | $\beta$ | $t$ | $p$ | $\beta$ | $t$ | $p$ | VIF |
| ENV (A) | 0.247 | 4.819 | .000 | 0.128 | 1.845 | .062 | 0.180 | 2.636 | .009 | 1.872 |
| GI (B) |  |  |  | 0.178 | 2.598 | .010 | -0.021 | -0.250 | .803 | 2.764 |
| Interaction (A × B) |  |  |  |  |  |  | -0.264 | -4.076 | .000 | 1.690 |
| $R^2$ (Adjusted $R^2$) | 0.061 (0.058) ||| 0.079 (0.073) ||| 0.120 (0.112) ||| |
| $F$ | 23.218 ($p<.001$) ||| 6.752 ($p<.05$) ||| 16.614 ($p<.001$) ||| |

*Note:* Dependent variable: SUS. ENV, environment; GI, green innovation; SUS, corporate sustainability

**Table 7|** The Moderating effect of green innovation (SOC)

|  | Model 1 ||| Model 2 ||| Model 3 ||| |
|---|---|---|---|---|---|---|---|---|---|---|
|  | $\beta$ | $t$ | $p$ | $\beta$ | $t$ | $p$ | $\beta$ | $t$ | $p$ | VIF |
| SOC (A) | 0.299 | 5.931 | .000 | 0.222 | 3.386 | .001 | 0.187 | 2.847 | .005 | 1.747 |
| GI (B) |  |  |  | 0.122 | 1.861 | 0.64 | 0.053 | 0.773 | .440 | 1.914 |
| Interaction (A × B) |  |  |  |  |  |  | -0.173 | -2.921 | .004 | 1.419 |
| $R^2$ (Adjusted $R^2$) | 0.090 (0.087) ||| 0.098 (0.093) ||| 0.120 (0.112) ||| |
| $F$ | 35.178 ($p<.001$) ||| 3.462 ($p>.05$) ||| 8.530 ($p<.05$) ||| |

*Note:* Dependent variable: SUS. SOC: social responsibility; GI, green innovation; SUS, corporate sustainability

**Table 8|** The Moderating effect of green innovation (GOV)

|  | Model 1 ||| Model 2 ||| Model 3 ||| |
|---|---|---|---|---|---|---|---|---|---|---|
|  | $\beta$ | $t$ | $p$ | $\beta$ | $t$ | $p$ | $\beta$ | $t$ | $p$ | VIF |
| GOV (A) | 0.273 | 5.363 | .000 | 0.213 | 4.066 | .000 | 0.202 | 3.849 | .000 | 1.113 |
| GI (B) |  |  |  | 0.199 | 3.806 | .000 | 0.163 | 2.930 | .004 | 1.254 |
| Interaction (A × B) |  |  |  |  |  |  | -0.100 | -1.839 | .067 | 1.197 |

| | | | |
|---|---|---|---|
| $R^2$ (Adjusted $R^2$) | 0.075 (0.072) | 0.111 (0.106) | 0.119 (0.112) |
| $F$ | 28.759 ($p<.001$) | 14.488 ($p<.05$) | 3.381 ($p>.05$) |

*Note:* Dependent variable: SUS. GOV: governance; GI, green innovation; SUS, corporate sustainability